# Temperature-driven massless Kane fermions in HgCdTe crystals: verification of universal velocity and rest-mass description

F. Teppe[1], M. Marcinkiewicz[1], S.S. Krishtopenko[1,2], S. Ruffenach[1], C. Consejo[1], A.M. Kadykov[1,2], W. Desrat[1], D. But[1], W. Knap[1], J. Ludwig[3,4], S. Moon[3,4], D. Smirnov[3], M. Orlita[5], Z. Jiang[6], S.V. Morozov[2], V.I. Gavrilenko[2], N.N. Mikhailov[7], S.A. Dvoretskii[7]

[1]*Laboratoire Charles Coulomb, UMR CNRS 5221, University of Montpellier, 34095 Montpellier, France*

[2]*Institute for Physics of Microstructures, Russian Academy of Sciences, 603950, GSP-105, Nizhny Novgorod, Russia*

[3] National High Magnetic Field Laboratory, Tallahassee, Florida 32310, USA

[4] Department of Physics, Florida State University, Tallahassee, Florida 32306, USA

[5] Laboratoire National des Champs Magnétiques Intenses, CNRS-UJF-UPS-INSA, 38042 Grenoble, France

[6] School of Physics, Georgia Institute of Technology, Atlanta, Georgia 30332, USA

[7] Institute of Semiconductor Physics, Siberian Branch, Russian Academy of Sciences, pr. Akademika Lavrent'eva 13, Novosibirsk, 630090 Russia

**Abstract**

It has recently been shown that the electronic states in bulk gapless HgCdTe offer another realization of pseudo-relativistic three-dimensional particles in a condensed matter system. These single valley relativistic states, referred to as massless Kane fermions, cannot be described by any other well-known relativistic massless particles. Furthermore, the HgCdTe band structure can be continuously tailored by modifying either the cadmium content or temperature. At the critical concentration or temperature, the bandgap, $E_g$, collapses as the system undergoes a semimetal-to-semiconductor topological phase transition between the inverted and normal alignments. Here, using far-infrared magneto-spectroscopy we explore the continuous evolution of band structure of bulk HgCdTe as temperature is tuned across the topological phase transition. We demonstrate that the rest-mass of the Dirac-like Kane fermions, $\widetilde{m}$, changes sign at the critical temperature, while their velocity, $\widetilde{c}$, remains constant. The relation $E_g = 2|\widetilde{m}|\widetilde{c}^2$ with the universal value of $\widetilde{c} = (1.07 \pm 0.05) \cdot 10^6$ m/s remains valid in a broad range of temperatures and Cd concentrations, indicating a striking universality of the pseudo-relativistic description of the Dirac-like Kane fermions in HgCdTe.

**Paper**

In condensed matter systems, the interaction of electrons with a periodic crystal lattice potential can give rise to low-energy quasiparticles that mimic the relativistic dynamics of Dirac particles in high energy physics. Perhaps the most spectacular demonstration of this concept was given ten years ago by the isolation of a monolayer of carbon atoms forming a graphene[1] sheet. The electrons in graphene behave as two dimensional (2D) massless fermions with gapless conical bands that obey the Dirac equation. Subsequently, further condensed matter analogues of high-energy relativistic fermions were demonstrated such as edge or surface states of 2D or 3D topological insulators[2-4] and 3D Dirac semimetals with linear energy-momentum dispersion in all three momentum directions[5-7].

Recently, another massless Dirac-like quasiparticle has been discovered in $Hg_{1-x}Cd_xTe$ at an inverted–to-normal band structure topological transition existing at the critical cadmium concentration[8] $x_C \approx 0.17$. These 3D massless Kane fermions are not equivalent to any other known relativistic particles. Similar to the pseudospin-1 Dirac-Weyl system[9], their energy dispersion relation features cones crossed at the vertex by an additional flat band. The band gap and the electronic dispersion in $Hg_{1-x}Cd_xTe$ can be tuned intrinsically by adjusting the chemical composition, or externally, by changing temperature[10]. The ability to control the properties of quasiparticles with relativistic behavior in a tabletop condensed-matter experiment holds vast scientific and technological potential. However, the variation of the chemical composition in $Hg_{1-x}Cd_xTe$ crystals does not allow for fine-tuning of the band gap in the vicinity of the phase transition due to inherent fluctuations of Cd concentration. Contrarily, a temperature-driven evolution of the band structure provides a conceptually straightforward, yet very accurate and detailed probe of the relativistic properties of Kane fermions while tuning the 3D $Hg_{1-x}Cd_xTe$ across the gapless state at the topological transition between a normal state and an inverted band gap state. The appearance of a non-zero gap does not exclude relativistic properties of the Kane fermions in $Hg_{1-x}Cd_xTe$, which is retained at energies significantly above the gap. The energy dispersion asymptotically tends to a linear behavior, as it is for relativistic electrons at high energies. The cut-off energies for relativistic behavior in the $Hg_{1-x}Cd_xTe$ compounds, as well as in other 3D Dirac-Weyl semimetals, are related to the presence of high-lying conduction bands and low-lying valence bands.

To describe the electronic structure near the center of the Brillouin zone in $Hg_{1-x}Cd_xTe$ close to $x_C$, we employ a simplified Kane model[11,12] taking into-account *k·p* interaction between the $\Gamma_6$ and $\Gamma_8$ bands, while neglecting the influence of the split-off $\Gamma_7$ band. The corresponding (6x6) Hamiltonian formally resembles the one for relativistic 3D Dirac particles (see Supplementary Materials). By neglecting small quadratic in momentum terms, the eigenvalues of this Hamiltonian can be presented in the form:

$$E_{\xi=0}(p) = 0, \quad E_{\xi=\pm 1}(p) = \widetilde{m}\tilde{c}^2 \pm \sqrt{\widetilde{m}^2\tilde{c}^4 + p^2\tilde{c}^2}, \qquad (1)$$

The first eigenvalue $\xi = 0$ corresponds to the flat heavy-hole band. The two other eigenvalues describe the electron ($\xi = +1$) and light-hole ($\xi = -1$) conical bands separated by $E_g = 2\widetilde{m}\tilde{c}^2$. This Dirac-like representation of Kane fermions in $Hg_{1-x}Cd_xTe$ contains only two parameters, the rest-mass $\widetilde{m}$ and the universal velocity $\tilde{c} = \sqrt{2P^2/(3\hbar^2)}$, while the material properties are introduced through the Kane's matrix element *P* and the $Hg_{1-x}Cd_xTe$ band gap $E_g$.

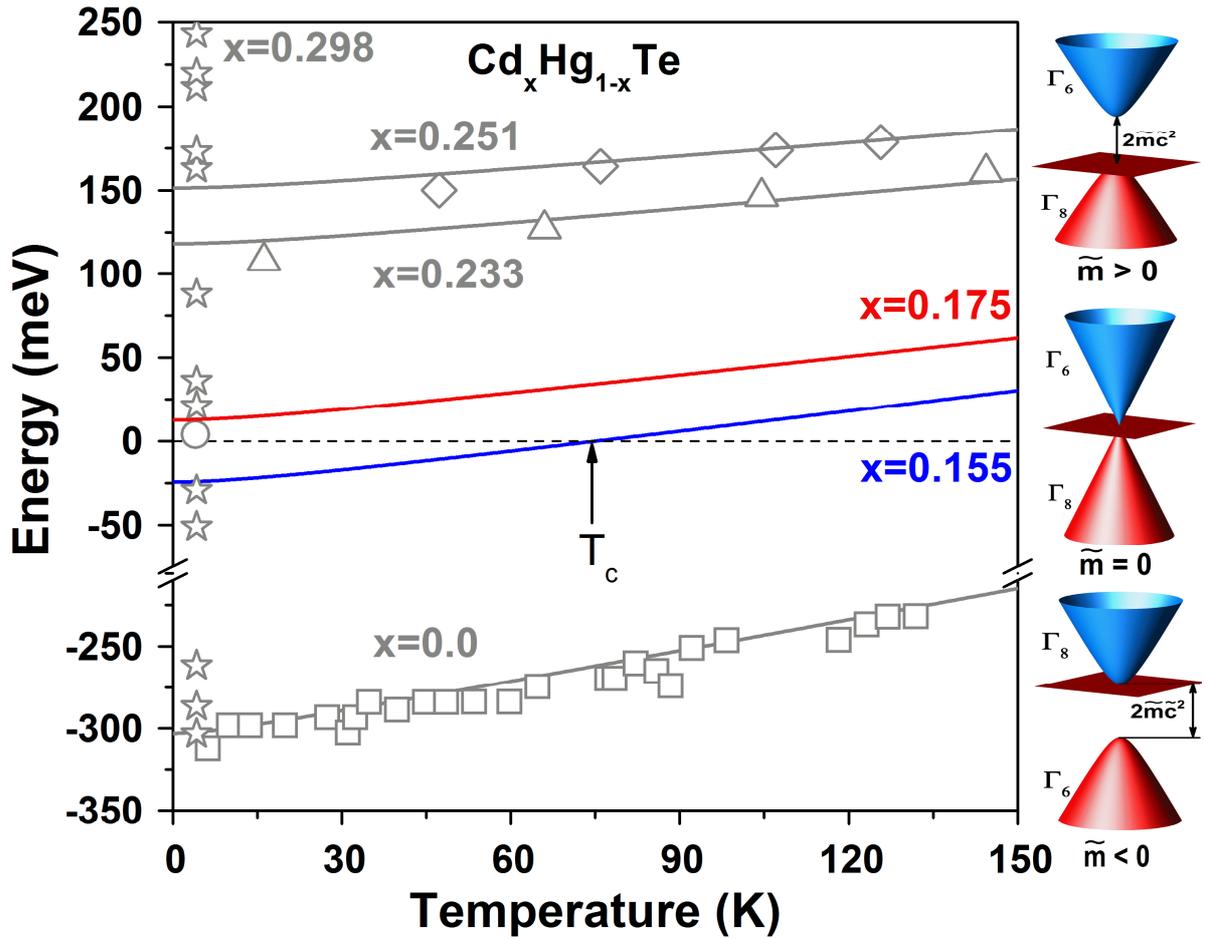

**Figure 1: Temperature tuning of the band structure in 3D $Hg_xCd_{1-x}Te$.**
The energy band gap in $Hg_xCd_{1-x}Te$, defined as the difference between the $\Gamma_6$ and $\Gamma_8$ band extrema at the center of the Brillouin zone, increases monotonously with temperature and Cd content. Open symbols represent experimental data from[8,13-15] for several Cd concentrations, and the lines show the $E_g(T)$ evolution calculated using equation 1 in Supplementary Material[15]. The red and blue lines correspond to the Cd concentration for samples studied in this work, $x = 0.155$ and $x = 0.175$. At the critical temperature, $T_C$, the system undergoes a semimetal-to-semiconductor topological phase transition between inverted and normal states. For $x = 0.155$, the gapless state is realized at $T_C \approx 77$ K. **Right: schematic band structure of Kane fermions in 3D $Hg_xCd_{1-x}Te$** As the rest-mass of Kane Fermions changes, the electronic dispersion evolves from a standard gapped semiconductor for $\widetilde{m} > 0$ into a semimetal at $\widetilde{m} < 0$. At the point of the topological transition, $\widetilde{m} = 0$, the conical conduction (blue) and the light-hole valence (red) bands are crossed at the vertex by a flat heavy-hole band (brown).

The evolution of Dirac-like Kane fermions in $Hg_{1-x}Cd_xTe$ is illustrated in Figure 1. If the rest-mass $\widetilde{m}$ is positive, the crystal is a typical narrow-gap semiconductor with the s-type $\Gamma_6$ band lying above the p-type $\Gamma_8$ bands, as schematically shown on the right side of Figure 1. On the other hand, if $\widetilde{m} < 0$, the band order is inverted: the $\Gamma_6$ band lies below the $\Gamma_8$ bands. Since the two $\Gamma_8$ bands always touch each other at the $\Gamma$ point of the Brillouin zone, the band structure is gapless and the crystal is a semimetal. The two distinct phases with different sign of the rest-mass are not topologically equivalent, as characterized by a $Z_2$ topological invariant[16].

Experimentally, the dispersion of gapless or gapped, 2D or 3D Dirac fermions can be conveniently probed through the magnetic field dependence of inter-Landau level transitions[17-23]. The application of a quantizing magnetic field $B$ transforms the zero-field continuum of states into a set of unequally spaced Landau levels (LLs) with a distinct $\sqrt{B}$ behavior. In pristine gapless graphene, for example, the LLs have a simple structure given by (after[17]): $E_n = \text{sgn}(n)\sqrt{2|n|e\hbar \tilde{c}^2 B}$, where $\hbar$ is the Planck's constant, and $e$ is the electron charge. The integer LL index $n$ labels electron- ($n > 0$) and hole-like ($n < 0$) states, and unconventional, zero-energy, field-independent $n = 0$ LL states. In graphene/boron-nitride heterostructures with zero crystallographic alignment angle, an intrinsic gap $\Delta$ opens up separating zeroth dispersionless LLs, $E_{n=0} = \pm\Delta/2$. Other $|n| > 0$ electron(hole)-like LLs shift up(down) by $\Delta/2$ as well: $E_{|n|>0} = \text{sgn}(n)\sqrt{2|n|e\hbar\tilde{c}^2 B + (\Delta/2)^2}$ (after[20]).

The LL spectrum of massive or massless Dirac-like fermions in $Hg_{1-x}Cd_xTe$ has a more complex form:

$$E_{\xi,n,\sigma}(p_z) = \xi^2 \tilde{m}\tilde{c}^2 + \xi\sqrt{\tilde{m}^2\tilde{c}^4 + \frac{e\hbar\tilde{c}^2 B}{2}(4n - 2 + \sigma) + \tilde{c}^2 p_z^2} \qquad (2)$$

Here, the LL index $n$ runs over positive integers $n = 1, 2, ....$ for the states in the electron and light-hole bands ($\xi = \pm 1$). For the zero energy, flat heavy-hole band ($\xi = 0$), $n$ runs over all nonnegative integers except 1: $n = 0, 2, 3 ....$ The quantum number $\sigma$ accounts for the Kramer's degeneracy lifted by the magnetic field $B = (0, 0, B_z)$. The corresponding splitting can be viewed as the Zeeman (spin) splitting of LLs: $E_{\xi,n,\uparrow}(p_z) - E_{\xi,n,\downarrow}(p_z)$. The non-parabolicity of the bands implies a strong dependence of this spin-splitting on $p_z$, $B$, and on LL and band indices.

Here, we conduct a systematic optical investigation of the dispersion of Dirac-like Kane fermions in $Hg_{1-x}Cd_xTe$ crystal as a function of temperature and magnetic field. From the experimental point of view, the use of temperature as a band structure tuning parameter in magneto-optical studies is challenging and demands ultimate quality samples. This is because the observation of well-defined optical resonances requires high carrier mobility, which degrades with increasing temperature due to the increase of scattering on phonons. Our bulk $Hg_{1-x}Cd_xTe$ samples were grown by molecular beam epitaxy. The Cd concentration was chosen to enable exploring the $E_g(x,T)$ space across the semimetal-to-semiconductor phase transition using temperature for fine "gap-at-will" tuning (Figure 1). The sample A, $x = 0.175$, is a standard narrow-gap semiconductor at any temperature. The sample B, $x = 0.155$, is a semimetal at low temperatures with a negative band gap corresponding to the inverted band order. As the temperature increases, the inverted band gap closes as the system undergoes a semimetal-to-semiconductor phase transition at the critical temperature $T_C \approx 77$ K followed by the opening of a gap in the normal state.

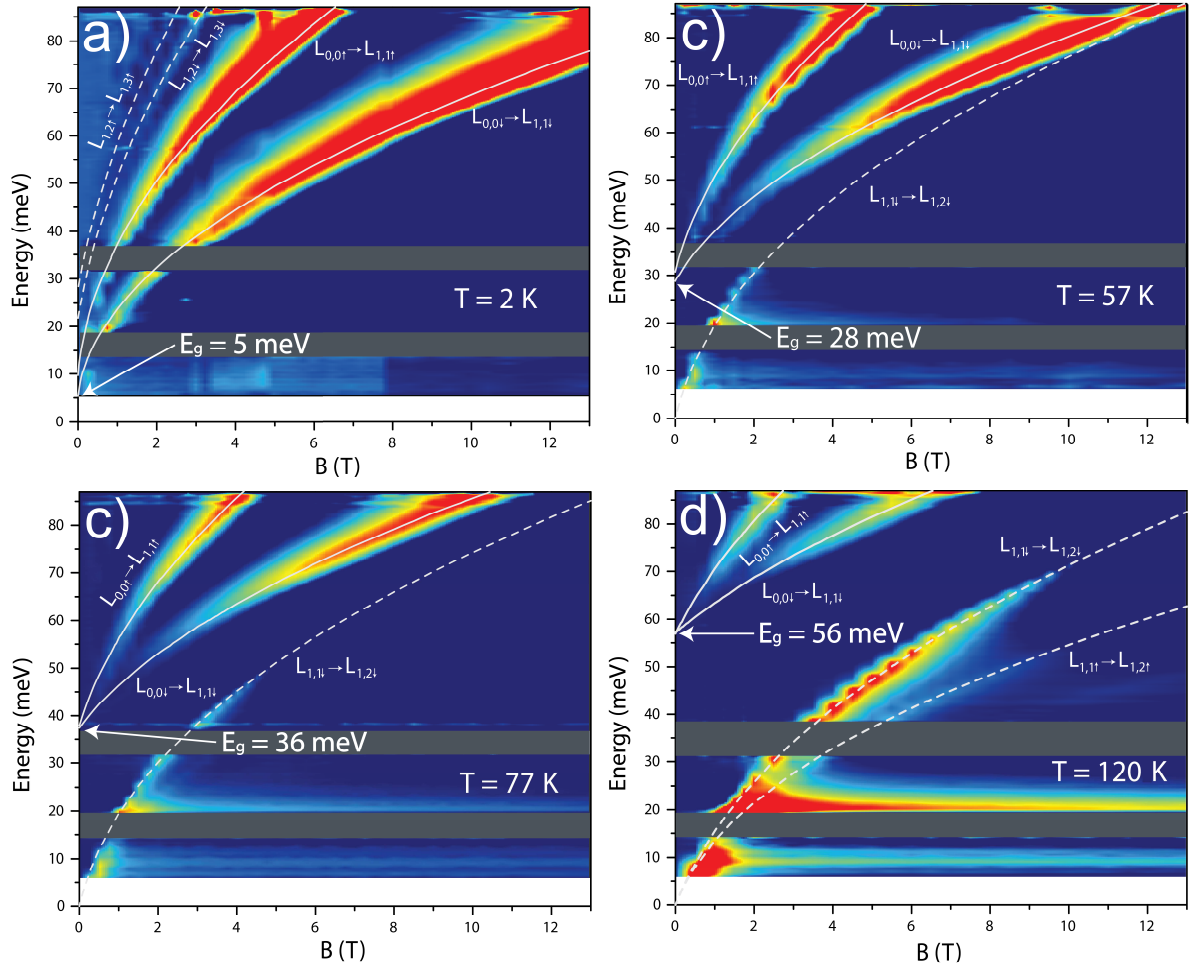

**Figure 2: Relative change of absorbance plotted as false color maps, exhibiting inter-Landau-level transitions as a function of magnetic field in the sample A at temperatures from 2 K to 120 K.** White lines represent the fits using the simplified Kane model, allowing for determination of $\tilde{m}$ and $\tilde{c}$. Solid lines correspond to inter-band transitions while dashed-lines account for intra-band transitions. The band-gap values at different temperatures, determined by the inter-band transition energies at zero magnetic field, are depicted with arrows. a) At 2 K, only inter-band transitions are seen and a $\sqrt{B}$-like behavior of inter-Landau-level resonances and spin splitting of Landau levels is observed. b) At 57 K intra-band transitions become visible in addition to previous lines. c) and d) At higher temperatures the energy difference between inter-band and intra-band lines at zero field clearly increases, corresponding to a gap opening as a function of temperature.

Besides the linear behavior of the absorption coefficient at zero magnetic field measured in both samples (see Figure 1 in Supplementary Materials), the presence of pseudo-relativistic 3D fermions with $\tilde{m} \approx 0$ is established at 2 K in the sample A by $\sqrt{B}$-like dependence of optical transitions and a spin splitting of LLs seen in Figure 2a. The fitting of the two main lines based on equation (2), shows that they are related to the inter-band transitions between the heavy-hole band remaining at zero energy ($\xi = 0$) and the $n = 1$ spin-up and spin down LLs with $\xi = 1$ (see for details Figure 2 in Suppl. Materials). The extracted band gap value at 2 K equals $2\tilde{m}\tilde{c}^2 = (5\pm2)$ meV. Magneto-optical results obtained at temperatures from 57 K to 120 K are shown in Figure 2b, 2c and 2d. The band-gap, visualized by intersect of the inter-band transitions with the energy axis (shown by white arrows), increases with temperature. In addition to inter-band transitions ($\Delta\xi = 1$) intra-band LLs transitions ($\Delta\xi = 0$) are also observed and fitted with equation (2).

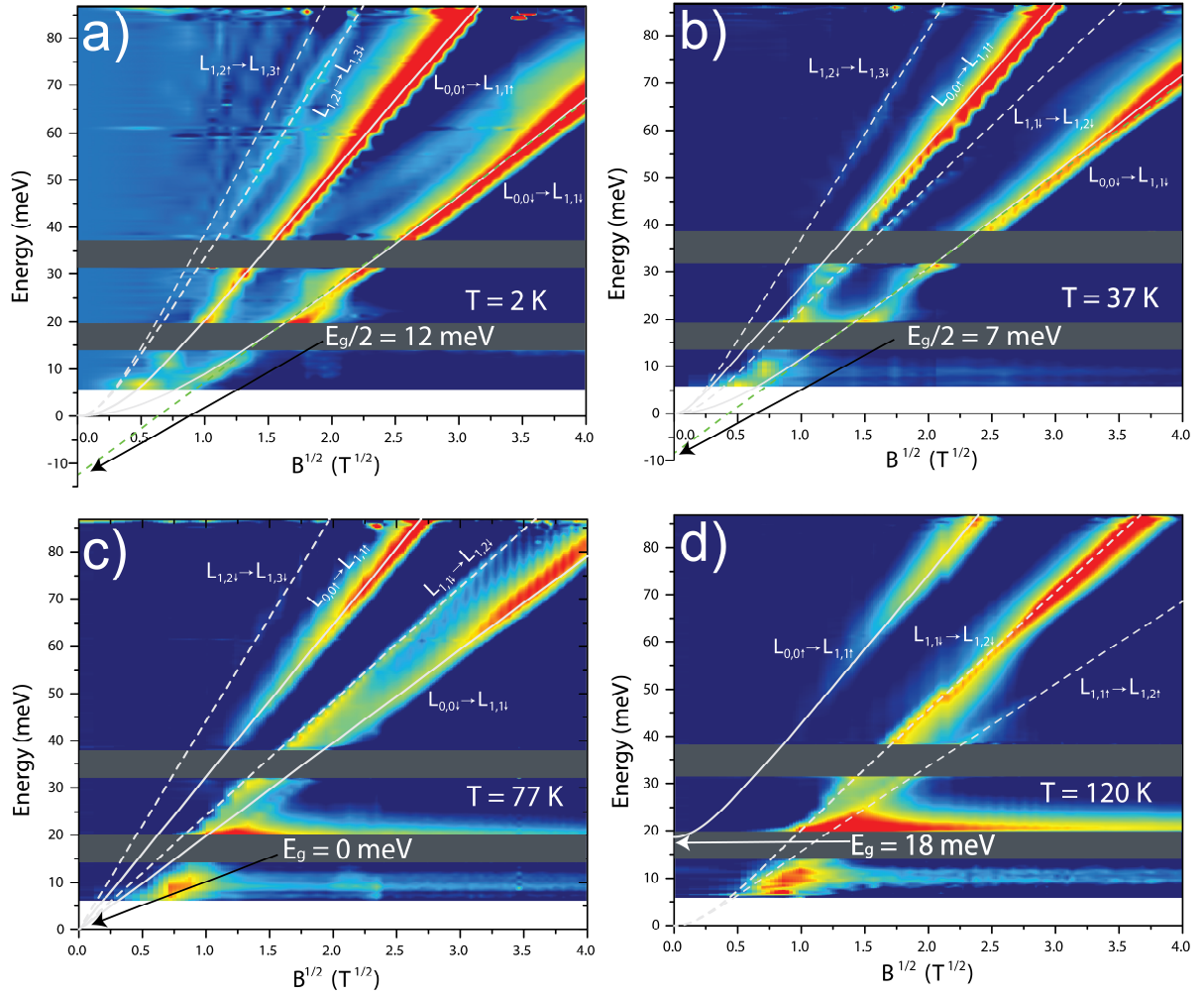

**Figure 3: Color maps of the inter-Landau-level transitions as a function of magnetic field in the sample B at temperatures from 2 K to 120 K in a square root scale.** White lines represent the fits using the simplified Kane model. Solid lines represent inter-band optical transitions while dashed lines correspond to intra-band transitions. Green dotted lines are guides for the eyes accounting for half the band-gap energy at each temperature (with arrows indicating half the band-gap values) **a)** and **b)** At 2 K and 37 K some discrepancies to the $\sqrt{B}$ behavior of inter-Landau-level optical transitions are observed, corresponding to the existence of a negative band gap. **c)** At 77 K a pure $\sqrt{B}$ behavior corresponds to the gapless state and the presence of genuine massless Kane fermions. **d)** At 120 K a positive gap opens as seen with the presence of an inter-band optical transition.

Magneto-absorption of the sample B at different temperatures is presented in Figure 3. in a $\sqrt{B}$-scale for a sake of clarity. It is seen that at low temperatures and low magnetic field values, the LLs transitions exhibit some discrepancies compared to a pure $\sqrt{B}$ behavior. Indeed, as shown in Figure 1, the linear energy dispersion in conduction and valence bands arises in gapless samples only. However, even in the case of small negative (or positive) gap, the bands could be considered as parabolic in the vicinity of the Γ point. It corresponds to the small values of parameter $p/\widetilde{m}\widetilde{c}$, for non-zero rest-mass values (see equation (4) in Supplementary Materials). At low magnetic fields, the band parabolicity results in a linear behavior of the LLs transitions as a function of *B*. Therefore, only a precise $\sqrt{B}$ behavior down to the lowest applied magnetic fields implies a system with genuine massless particles. At temperatures below $T_c$, the deviation from a pure $\sqrt{B}$ behavior is well reproduced by the theory and gives absolute values of the rest-mass $\widetilde{m}$ approaching to zero when temperature increases. As seen in equation (2), linear extrapolation of optical transitions in the square-root scale intersects the energy axis at $E = E_g/2$, as represented by arrows in Figure 3a and 3b. An accurate $\sqrt{B}$ behavior for all optical transitions is obtained as an evident proof of gap closing at 77 K (Figure 3c). Above the critical temperature, the difference in energies of inter- and intra-band transitions in low magnetic fields becomes visible, as it is for the sample A, meaning that a positive gap between the $\Gamma_6$ and $\Gamma_8$ bands opens. This allows us claiming that at $T_c$ = 77 K a temperature-driven topological phase transition with pseudo-relativistic massless Kane fermions occurs. This fact is highlighted by the change of a sign of particle rest-mass seen in Figure 4a.

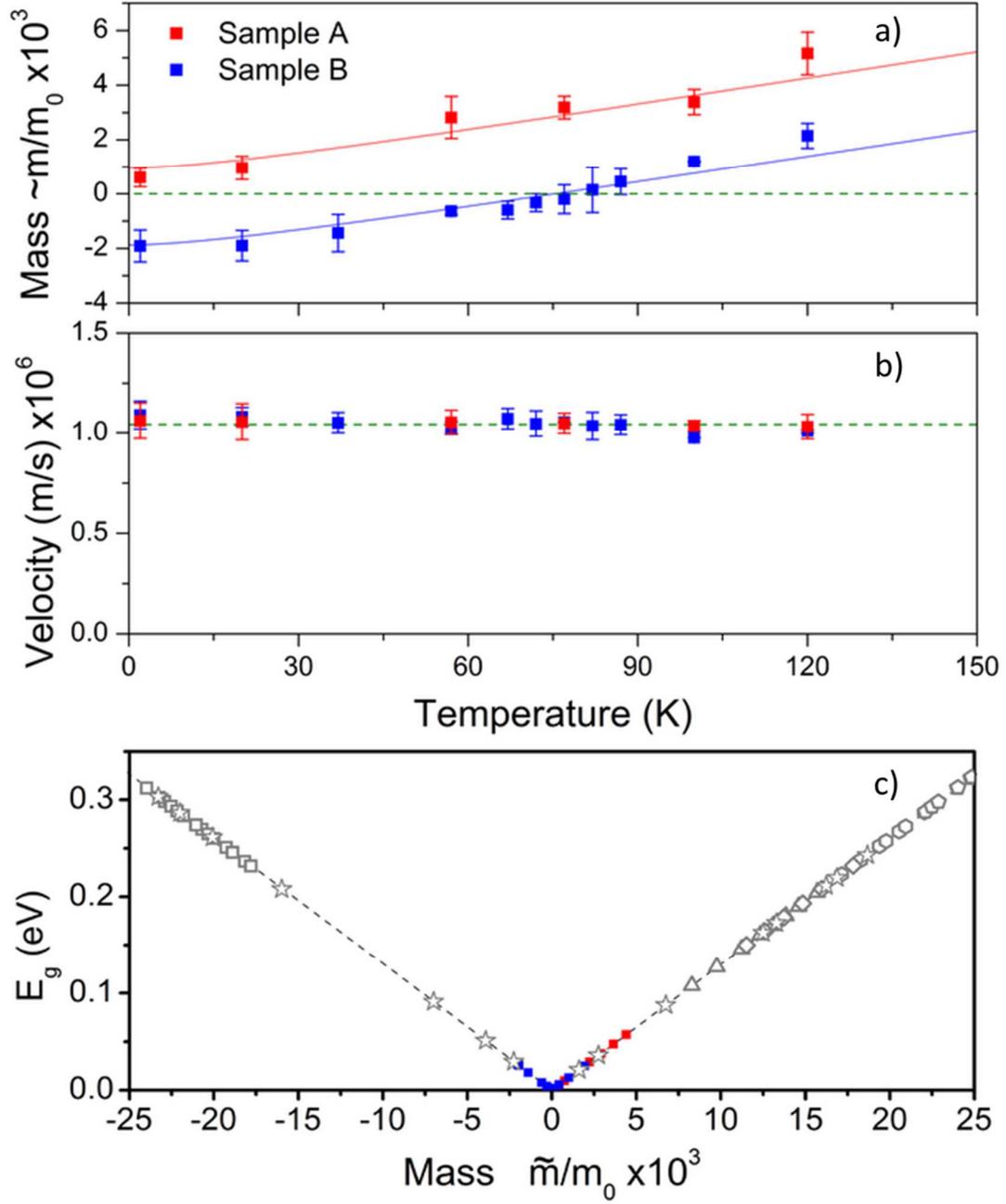

**Figure 4: Kane fermion parameters.** Blue and red points correspond to experimental results obtained in this work, while open symbols are experimental points from[13-15]. **a)** Rest-mass values for both samples are determined by fitting magneto-optical data with the simplified model, including the $\Gamma_6$ and $\Gamma_8$ bands only. Dirac-like Kane fermion rest-mass in the sample A is close to zero at low temperatures and increases with temperature up to 120 K. In the sample B, the Dirac-like particle rest-mass is negative at low temperatures, it vanishes at 77 K and then becomes positive above $T_c$. The lines are theoretical curves calculated using equation 1 in Supplementary Material [15], as seen in Figure 1. **b)** The velocity is nearly constant in both samples between 2 K and 120 K. The green dashed line represents the theoretical value. **c)** The universal evolution of the bandgap energy versus the Dirac-like Kane fermion rest-mass in HgCdTe alloys is shown. The black dashed line serves as a guide for the eyes.

The rest-mass extracted from magneto-optical data and its dependence on temperature is compared with theoretical values using equation (7) in[15]. In the sample A, the rest-mass is positive and increases in the whole range of temperatures, as it is shown in Figure 4a. As discussed above, in the sample B the Kane fermion rest-mass experiences a change of sign corresponding to the temperature induced semimetal-to-semiconductor topological phase transition, which occurs at 77 K. Equation (7) in[15] describes very well the experimental rest-mass curves for both samples and clearly reproduces the phase transition at 77 K in the sample B. The conic dispersion relation of the massless particles can be therefore realized for the specific range of crystal chemical composition and an according temperature - conditions which provide that the bulk band gap is fully closed. Interestingly, the Dirac-like Kane fermion velocity $\tilde{c}$ is nearly constant over the whole range of temperatures for both samples with Cd contents of 0.155 and 0.175. The extracted value of $\tilde{c} = (1.07 \pm 0.05) \cdot 10^6$ m/s is in a very good agreement with the theoretical value defined by $\tilde{c} = \sqrt{2P^2/(3\hbar^2)}$, which equals to $1.05 \cdot 10^6$ m/s for the well-accepted value of $E_P = 2m_0P^2/\hbar^2 \approx 18.8$ eV[24]. Therefore, this universal value of $\tilde{c}$ allows for determination of the particle rest-mass for bandgap values in the vicinity of the semimetal-to-semiconductor phase transition induced by temperature, Cd content, or other external parameter (e.g. pressure). Figure 4c shows the variation of the experimental bandgap energy obtained in this work (full blue and red points) and from previous studies[14,15] (open symbols), as a function of the rest-mass, using $E_g = 2\tilde{m}\tilde{c}^2$ and the universal value of $\tilde{c}$.

There are two points limiting the applicability of the simplified Kane model, considering the $\Gamma_6$ and $\Gamma_8$ band only, for actual HgCdTe crystals. The first one, already mentioned above, is related to the existence of other bands, considered as remote and not included in the model. The energy gap between the second and the lowest conduction bands in CdTe exceeds 4 eV, while the corresponding gap in HgTe is about 3 eV[25]. Therefore, the cut-off energies for conduction bands in the simplified model should be lower than 3 eV. For the valence band, the cut-off energy is defined by the energy difference $\Delta \approx 1$ eV between the split-off $\Gamma_7$ band, and the heavy-hole band. The second limitation is attributed to the flat heavy-hole band, characterized by an infinite effective mass in the model. To ignore the parabolic terms in the electron dispersion of the heavy-hole band, one has to consider sufficiently low energies $E$, such that the relativistic mass of the fermions $E/\tilde{c}^2$ should be significantly lower than the heavy-hole mass $m_{hh}$. Assuming $m_{hh} \approx 0.5m_0$, where $m_0$ is the free electron mass, we arrive at a cut-off energy of about 3 eV for the flat band approximation, which exceeds $\Delta$.

In conclusion, by using temperature as a fine-tuning external parameter we measured the bandgap energy of HgCdTe bulk crystals with well-chosen chemical composition in the vicinity of the semimetal-to-semiconductor phase transition. We observed genuine massless Kane fermions at the critical temperature of 77 K. We used the simplified Kane model to determine the pseudo-relativistic Dirac-like Kane fermion parameters $\tilde{m}$ and $\tilde{c}$ as a function of temperature and Cd content. We observed a change of sign of $\tilde{m}$ accounting for the temperature-driven topological phase transition. Our results also reveal universal velocity in HgCdTe crystals allowing for the determination of the Dirac-like Kane fermion rest-mass from all experimental results in the literature obtained in the distant vicinity of the phase transition.

**Methods**

We performed magneto-optical studies on two [013]-oriented $Hg_{1-x}Cd_xTe$ layers, with different Cd concentrations, x = 0.17 and 0.155. Both films were sufficiently thick (≈3.2 µm) to be considered as 3D materials, and thin enough to be transparent in the far-infrared (FIR) spectral range. The samples were grown by molecular beam epitaxy (MBE) on semi-insulating GaAs substrates with relaxed CdTe buffers[26]. We used a special ultra-high vacuum multi-chamber MBE set which allows for the growth of very high quality HgCdTe crystals monitored by in situ reflection high-energy electron diffraction and single wavelength ultra-fast ellipsometry (0.5 nm).

The magneto-optical transmission measurements were carried out by using a Fourier transform spectrometer coupled to a 16 Tesla superconducting coil. The radiation of a Globar lamp is guided to a sample using oversized waveguides (light pipes). The intensity of the transmitted light is measured by a silicon bolometer. Both the magnet and the bolometer require cryogenic temperatures, therefore most of reported up to today experiments were conducted at 4.2 K or lower temperatures. In this work, in order to perform a temperature tuning of the band structure, the standard magneto-optical configuration required important modifications. The bolometer was placed in a vacuum chamber separated from the sample chamber. To provide a wide spectral range for experiments, an indium sealed cold diamond window ensures the optical coupling between the transmitted light and the bolometer. An additional superconducting coil around the bolometer compensates the spread field of the main coil, keeping the bolometer at zero magnetic field. This additional compensating superconducting coil also provides an additional screening of the bolometer. A Lambda plate coil – placed below the main magnet allows to obtain superfluid helium around the bolometer and keep the main coil at 4 K. Therefore, this modified experimental set-up allows to keep the coils in their superconducting state, the bolometer at its optimal temperature, and to tune the sample chamber temperature in the 2 K – 150 K range.

The magneto-optical spectra were measured in the Faraday configuration up to 16 T, with a spectral resolution of 0.5 meV. All the spectra were normalized by the sample transmission response at $B = 0$ T.

**Acknowledgements**
The authors acknowledge M. Dyakonov, W. Zawadzki and A. Raymond for helpful discussions. This work was supported by the CNRS through LIA TeraMIR project, by the Languedoc-Roussillon region via the Terahertz Gepeto platform, by European cooperation through the COST action MP1204, and the Era.Net-Rus Plus project "Terasens". This work was also supported by the Russian Academy of Sciences, the non-profit Dynasty foundation, the Russian Foundation for Basic Research (Grants 14-02-31588, 15-02-08274, 16-02-00672), by Russian Ministry of Education and Science (Grant Nos. MK-6830.2015.2, HIII-1214.2014.2) and by grant RBFR 15-52-16017 NTSIL. J.L., S.M., Z.J. and D.S. acknowledge the support from the U.S. Department of Energy (Grant No. DE-FG02-07ER46451) for IR spectroscopy measurements at 4.2K that were performed at the National High Magnetic Field Laboratory, which is supported by National Science Foundation (NSF) Cooperative Agreement N° DMR-1157490 and the State of Florida.

## Supplementary materials

The band-gap energy versus temperature is given by the empirical formula found in[15]:

$$E_g(eV) = -0.303(1-x) + 1.606x - 0.132x(1-x) + \frac{[6.3(1-x) - 3.25x - 5.92x(1-x)]10^{-4}T^2}{[11(1-x) + 78.7x + T]} \quad (1)$$

Relativistic fermions are usually described by the Dirac equation:

$$i\hbar \frac{\partial \Psi}{\partial t} = \left(\beta mc^2 + c\alpha_x p_x + c\alpha_y p_y + c\alpha_z p_z\right)\Psi, \quad (2)$$

where $p_i$ ($i = x, y, z$) are the components of momentum operator, $m$ and $c$ are the rest mass and velocity of light respectively. The matrices $\alpha_i$ and $\beta$ define the symmetry properties of the particles and have the form

$$\beta = \begin{pmatrix} I & 0 \\ 0 & -I \end{pmatrix}, \qquad \alpha_i = \begin{pmatrix} 0 & \sigma_i \\ \sigma_i & 0 \end{pmatrix}, \quad (3)$$

in which $\sigma_i$ are the Pauli matrices and $I$ is a 2x2 unit matrix. As it is clearly seen from equation (1), if the rest-mass of particles equals zero, their dispersion is described by two-fold degenerate cone in energy-momentum space.

Current physics has proven the existence of several bulk condensed-matter materials which are fairly well described by the above equation. At the same time, there are also other systems with relativistic-like charge carriers, nevertheless, they are described by different Hamiltonians. For Kane fermions, the corresponding Hamiltonian $\hat{H}$ formally resembles the one for genuine 3D Dirac particles. However, we see that it has a form of a 6x6 matrix, which describes qualitatively a different system:

$$\hat{H} = \tilde{\beta}\tilde{m}\tilde{c}^2 + \tilde{c}\tilde{\alpha}_x p_x + \tilde{c}\tilde{\alpha}_y p_y + \tilde{c}\tilde{\alpha}_z p_z, \quad (4)$$

where $\tilde{\beta} = \begin{pmatrix} U_c & 0 \\ 0 & U_c \end{pmatrix}$, $\tilde{\alpha}_x = \begin{pmatrix} J_x & 0 \\ 0_i & -J_x \end{pmatrix}$, $\tilde{\alpha}_y = \begin{pmatrix} J_y & 0 \\ 0 & J_y \end{pmatrix}$, $\tilde{\alpha}_z = \begin{pmatrix} 0 & J_z \\ J_z & 0 \end{pmatrix}$,

and where $U_c$, $J_x$, $J_y$, $J_z$ are 3x3 matrices described as follows:

$$U_c = \begin{pmatrix} 1 & 0 & 0 \\ 0 & -1 & 0 \\ 0 & 0 & -1 \end{pmatrix}, \quad J_x = \begin{pmatrix} 0 & \frac{\sqrt{3}}{2} & -\frac{1}{2} \\ \frac{\sqrt{3}}{2} & 0 & 0 \\ -\frac{1}{2} & 0 & 0 \end{pmatrix}, \quad J_y = \begin{pmatrix} 0 & \frac{i\sqrt{3}}{2} & \frac{i}{2} \\ \frac{i\sqrt{3}}{2} & 0 & 0 \\ \frac{i}{2} & 0 & 0 \end{pmatrix}, \quad J_y = \begin{pmatrix} 0 & 0 & -1 \\ 0 & 0 & 0 \\ -1 & 0 & 0 \end{pmatrix}$$

Here we deliberately present low-energy Hamiltonian, which describes the Kane fermions in a form similar to the Dirac equation, defining the rest-mass $\tilde{m}$ and velocity $\tilde{c}$ of the Kane fermions. The matrices $J_x$, $J_y$, $J_z$ arising in equation (1) do not satisfy the algebra of angular momentum 1, therefore the Hamiltonian $\hat{H}$ does not reduce to any well-known case of relativistic particles. However, the Kane fermions, described by equation (4), share a number of properties with other relativistic particles.

Due to the conical dispersion of the massless Kane fermions in HgCdTe crystals, absorption coefficient $\lambda$ of both samples measured at 4.2 K exhibits a linear behavior as a function of photon energy $\hbar\omega$ at high energies similar to 3D Weyl systems[9]. This linear behavior of the optical conductivity provides a clear evidence for the pseudo-relativistic behavior of the 3D particles in any Dirac/Weyl-like materials, as discussed by Timusk[27]. However, if the HgCdTe layer is not exactly gapless, the hyperbolic dispersions mimic the linear energy-momentum law at higher energies, exceeding the band gap values. The relatively simple form of eigenvalues of the Kane fermion's Hamiltonian in equation (1) allows one to calculate the absorption coefficient above $2\widetilde{m}\widetilde{c}^2$ as a function of $\hbar\omega$ analytically[28]:

$$\lambda(\hbar\omega) = \frac{K}{(\hbar\omega)^3}\left[(1-\eta)\sqrt{(1-\eta)^2-\eta^2} + \frac{1}{8}\sqrt{1-4\eta^2}\right], \quad (5)$$

where $\eta = \widetilde{m}\widetilde{c}^2/\hbar\omega$ and $K$ is a constant. The first term in brackets describes absorption coefficient involving flat (heavy-hole) band, while the second term corresponds to the contribution of the light hole band. By analyzing the absorption coefficient with equation (2) at energies exceeding the GaAs substrate Reststrahlen band, shown in the figure below, we extract the band gap values $2\widetilde{m}\widetilde{c}^2$ at low temperatures for both samples, which are approximately $4 \pm 2$ meV and $-20 \pm 4$ meV for the samples A and B, respectively. As it is for Dirac fermions obeying equation (1), the appearance of non-zero band gap does not exclude pseudo-relativistic character of the Kane fermions, which is retained at energies significantly above (or below) $2\widetilde{m}\widetilde{c}^2$. In our case, the presence of pseudo-relativistic 3D fermions is also revealed by the linear character of absorption coefficient as a function of $\hbar\omega$ at high energies.

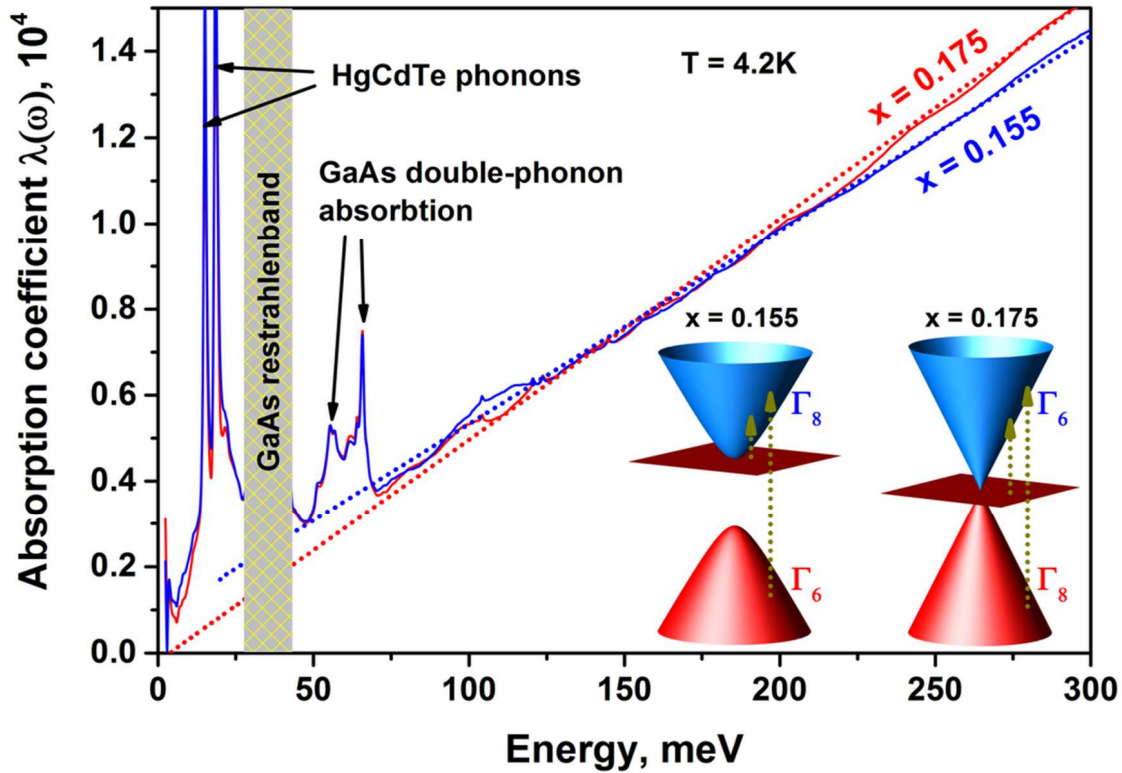

**Figure 1: Optical absorption of pseudo-relativistic Kane fermions in Hg$_{1-x}$Cd$_x$Te.** Zero field absorption coefficients exhibit a linear behavior reflecting the relativistic character of the 3D Kane fermions in Hg$_{1-x}$Cd$_x$Te. The band gap values of $2\widetilde{m}\widetilde{c}^2 = (4\pm2)$ and $(20\pm4)$ meV for $x = 0.175$ and $x = 0.155$, respectively, are extracted from fits (dashed lines) based on equation (2). The inset cartoon depicts inter-band transitions that contribute to the linear optical absorption.

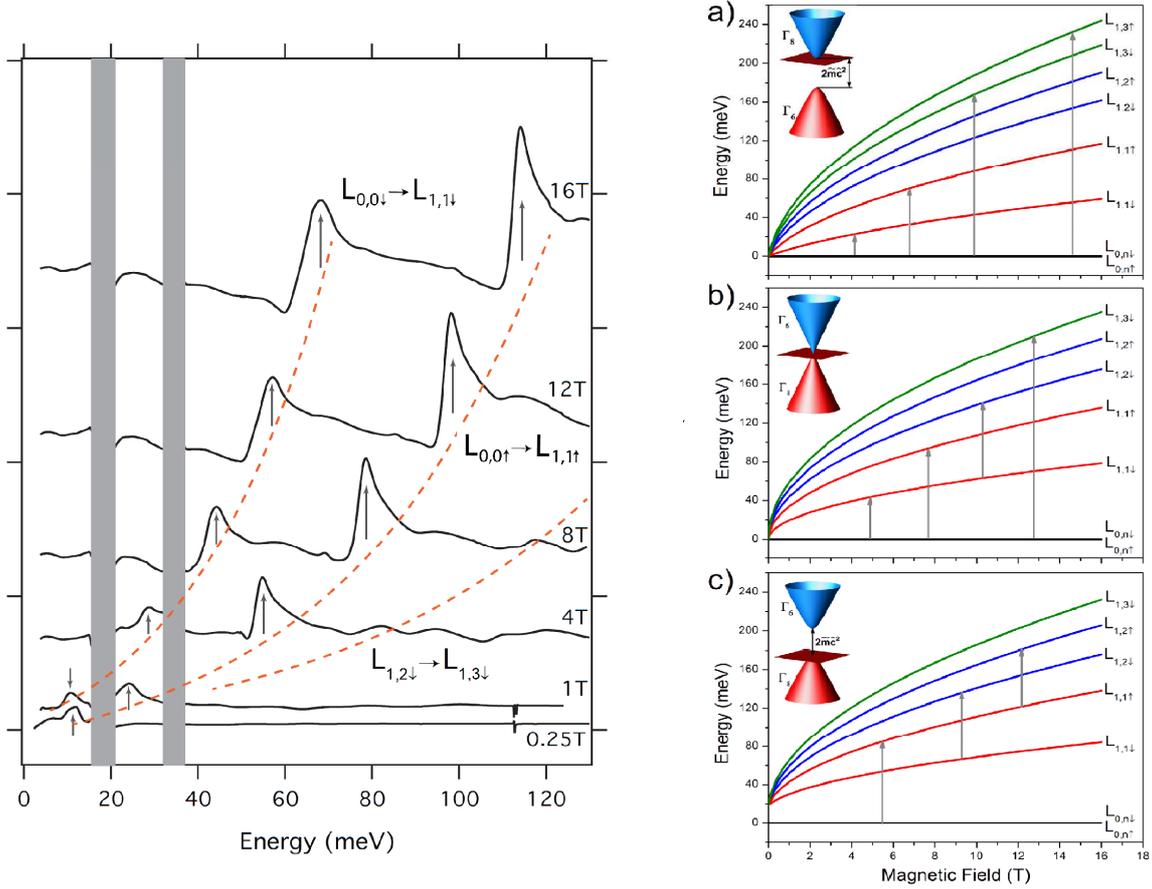

**Figure 2: Experimentally measured magneto-absorption spectra (right side).** Examples of transmission spectra measured at 4 K in sample A are shown for different magnetic field values up to 16 T. Two asymmetrical absorption lines corresponding to inter-band transitions are clearly observed at energies of the optical transitions denoted by arrows in the right panel (Figure b). Other symmetrical (intra-band) lines are also observed at higher energies. **Landau-level fanchart (left side).** Landau levels in a) inverted, b) gapless, and c) normal band structure HgCdTe compounds as a function of magnetic field, calculated using the simplified Kane model for $x = 0.155$ and T = 2 K, 77 K, and 120 K, respectively. Arrows show the optically allowed transitions observed in our magneto-transmission spectra.